\documentclass[journal=jacsat,manuscript=article]{achemso}

\usepackage[version=3]{mhchem} 
\usepackage[ruled,vlined]{algorithm2e}
\usepackage{graphicx}
\usepackage{subcaption} 
\usepackage{overpic} 
\usepackage{xcolor}

\newcommand{\beginsupplement}{%
        \setcounter{table}{0}
        \renewcommand{\thetable}{S\arabic{table}}%
        \setcounter{figure}{0}
        \renewcommand{\thefigure}{S\arabic{figure}}%
     }

\author{Luca Murg}
\affiliation{Theoretical Division, Los Alamos National Laboratory, Los Alamos, New Mexico 87545, USA}
\email{rtutchton@lanl.gov}
\alsoaffiliation{Department of Nuclear, Plasma, and Radiological Engineering, University of Illinois at Urbana-Champaign, Urbana, USA}
\author{Christopher Lane}
\affiliation{Theoretical Division, Los Alamos National Laboratory, Los Alamos, New Mexico 87545, USA}
\author{Roxanne M. Tutchton}
\affiliation{Theoretical Division, Los Alamos National Laboratory, Los Alamos, New Mexico 87545, USA}

\title[An \textsf{achemso} demo]
  {Gaunt and Breit Two-electron contributions to Mean-field Transformations and Fine Structure Splitting}

\abbreviations{IR,NMR,UV}

\begin{document}

\begin{abstract}
Materials utilized by novel energy systems are often studied using weakly correlated mean-field theories. However, if these systems incorporate heavy elements, relativistic effects must be included. Therefore a Kramers unrestricted Coupled Cluster with singles and doubles excitation formalism within a molecular mean-field Exact Two-Component framework (X2C$_{mmf}$) using a four-component Dirac-Hartree-Fock (DHF) reference state is presented. The exact X2C$_{mmf}$ transformed normal-order Hamiltonian incorporates all one-electron and two-electron (2e) contributions from the Coulomb, Gaunt, and Breit operators and is used with the Equation of Motion method to calculate the excitation energies of the alkali group of elements. Using this framework, the effects of 2e Gaunt and Breit integrals are studied. Results demonstrate growing contributions from these integrals to the generated X2C$_{mmf}$ mean-fields and electronic fine structure calculations with increasing atomic number. Overall, this paper outlines the method and effect of a higher level of accuracy within the X2C$_{mmf}$ approach and lays the foundation for future theoretical development of relativistic calculations within this framework.
\end{abstract}

\section{Introduction}
Materials utilized by novel energy systems, e.g., Generation IV nuclear reactors\cite{murg2025enhanced}, are often studied using weakly correlated mean-field theories. In weakly correlated electron systems, the direct Coulomb interaction can be modeled by an effective single-particle potential, e.g., Density Functional Theory. When strong correlations dominate, these mean-field theories can be used as a reference state for various strongly correlated approaches\cite{himmetoglu2014hubbard,georges1996dynamical,wodynski2022local,rivero2009description}. Relativistic effects are typically added perturbatively\cite{hess1996mean,schwerdtfeger2002relativistic,klein2008perturbative}, but this framework is not adequate for systems composed of heavy elements\cite{tutchton2020electronic,pyykko1988relativistic,saue2011relativistic,autschbach2012perspective,pyykko2012relativistic,liu2014advances,jayatilaka1992form} and the vast unexplored space of  topological materials\cite{fukui2007topological,rau2016spin}. These deficiencies require us to adjust existing theories to account for a four-component (4c) Dirac reference state\cite{swirles1935relativistic} utilizing increasingly accurate relativistic Hamiltonians, e.g., the Dirac-Coulomb (DC), Dirac-Coulomb-Gaunt (DCG), and  Dirac-Coulomb-Breit (DCB) Hamiltonians. In the no-pair approximation, Exact Two-Component (X2C) transformation theory can be used to decouple the positive energy (pe) and negative energy (ne) spectrum of the Dirac Hamiltonian for wave function-based correlation methods\cite{liu2018two,asthana2019exact,pototschnig2021implementation,tecmer2014communication,sikkema2009molecular}. A variety of X2C flavors have been developed most of which fall under the one-electron (1e) X2C framework\cite{peng2012exact,knecht2022exact}. In the case of the 1eX2C Hamiltonian scheme, the two-electron (2e) interaction terms are omitted from the defining 4c Dirac Hamiltonian resulting in a two-component (2c) Hamiltonian that is “exact” only in the 1e terms. Because the 2e interaction terms are often left untransformed in the 1eX2C basis set, 2e picture-change effects (2ePCEs) will appear.\cite{knecht2022exact}. In contrast to the 1eX2C transformation, performed before the self-consistent field (SCF) cycle, molecular mean-field X2C (X2C$_{mmf}$)  transformations utilize a unitary decoupling of the 4c molecular mean-field Fock matrix after having converged the 4c SCF DHF equations\cite{sikkema2009molecular}. Formally, X2C$_{mmf}$ transformation are exact. However, due to the computational cost associated with recomputing and transforming 2e integrals associated with Coulomb ($G^{LL}_{LL}$, $G^{LS}_{LS}$, $G^{SL}_{SL}$, $G^{SS}_{SS}$) and Gaunt/Breit integrals  ($G^{SS}_{LL}$, $G^{SL}_{LS}$, $G^{LS}_{SL}$, $G^{LL}_{SS}$), previous work using X2C$_{mmf}$ transformations have largely only considered the large component Coulomb integrals in the 2e terms of the X2C$_{mmf}$ transformed normal-order Hamiltonian and often in untransformed form ($G^{++}_{++} \approx G^{LL}_{LL}$). Neglecting or using untransformed 2e interaction terms results in only an approximate X2C$_{mmf}$ transformation\cite{saue2020dirac,knecht2022exact,sikkema2009molecular,zhang2024dirac}.\\

This work demonstrates that Gaunt and Breit 2e integral's contribution to the generated mean-field grows as elements become heavier. Furthermore, variations of the X2C$_{mmf}$ transformations including the exact effective 1e DC, DCG, and DCB Fock matrix in-conjunction with the approximate and exact 2e transformed Coulomb, Coulomb-Gaunt, and Coulomb-Breit interaction\cite{sikkema2009molecular} are computed. Using these X2C$_{mmf}$ transformed normal-order Hamiltonians, Kramers unrestricted Coupled Cluster (CC) with singles and doubles (SD) excitation formalism is applied\cite{vcivzek1966correlation,vcivzek1969use,vcivzek1971correlation,bartlett2007coupled}. The X2C$_{mmf}$-CCSD ground state wave function is used as reference for the Equation of Motion (EOM) method to obtain the excited wave function of the alkali elements\cite{saue2020dirac,yuwono2025two,stanton1993equation}. Results offer insights into the importance of the Gaunt and Breit 2e integrals contributions to the 2e terms of the exact X2C$_{mmf}$ transformed normal-order Hamiltonian used in post 4c SCF correlation-excitations steps as elements become heavier.

\section{Theory and Methodology}
In this work, PySCF's\cite{sun2018pyscf,sun2020recent} implementation of 4c DHF\cite{sun2021efficient,sun2022efficient}using the DC, DCG, and DCB Hamiltonian was used (see Fock matrix in Equation \ref{eq:fock}). The 4c DHF calculation enforced the restricted kinetic balance condition\cite{sun2021efficient,sun2022efficient} and used the finite-size nuclear model (convergence tolerance set to 1e$^{-10}$, max direct inversion of the iterative subspace\cite{csaszar1984geometry} (DIIS) set to 8).

\begin{equation}
\begin{split}
    \begin{pmatrix} F^{LL} & F^{LS}\\ F^{SL}  & F^{SS}\end{pmatrix} \begin{pmatrix} C^{+}_{L} & C^{-}_{L}\\ C^{+}_{S} & C^{-}_{S}\end{pmatrix} = \begin{pmatrix} I \otimes S & 0\\ 0 & \frac{1}{2mc^2}I \otimes T\end{pmatrix} \\
    \begin{pmatrix} C^{+}_{L} & C^{-}_{L}\\ C^{+}_{S} & C^{-}_{S}\end{pmatrix} \begin{pmatrix} \epsilon^{+} & 0\\ 0 & \epsilon^{-}\end{pmatrix},
    \label{eq:fock}
\end{split}
\end{equation}

\noindent A code of Kramers unrestricted CCSD (using NumPy) \cite{visscher1995kramers,visscher1996formulation} was created employing DIIS following standard CCSD methods for post-SCF wavefunction correlation calculations. The family of relativistic ANO basis sets\cite{roos2004main, roos2005new} used for this work are obtained from Basis Set Exchange\cite{pritchard2019new}. We note that in 4c DHF, the 2e integral $\hat{g}$ operator takes the following form:

\begin{center}
    \textbf{\textcolor{blue}{\hspace{10mm}\text{Breit}}}
\end{center}
\begin{equation}
\hat{g}_{ij} =  \frac{I_4}{r_{ij}} -\frac{1}{2} \left( \frac{\alpha_i \cdot \alpha_j}{r_{ij}} + \frac{\alpha_i \cdot r_{ij} \alpha_j \cdot r_{ij}}{r_{ij}^3} \right).
\label{eq:eri_DCB}
\end{equation}

\begin{center}
    \textbf{\textcolor{blue}{\hspace{-7mm}\text{Coulomb} \hspace{1mm} \text{Gaunt}\hspace{8mm} \text{gauge}}}
\end{center}

\noindent The first term in this expression is identified as Coulomb, the second as Gaunt, and the third as a gauge. The second and third term are often collectively called the Breit term and are associated with spin-orbit interactions and retardation effects. As dictated by an X2C$_{mmf}$ transformation, after converging the 4c DHF equation, the 
Hamiltonian for the valence electrons can be written in normal-ordered form,

\begin{equation}
\hat{H}_{X2C_{mmf}} = \sum_{PQ} F^Q_P \, \{a^\dagger_P a_Q\} + \frac{1}{4} \sum_{PQRS} V^{QS}_{PR} \, \{a^\dagger_P a^\dagger_R a_S a_Q\},
\end{equation}
for further use in wave function-based correlation methods\cite{sikkema2009molecular,pototschnig2021implementation}. The brackets indicate normal ordering with respect to occupied (hole) and virtual (particle) orbitals.   The effective 1e terms of the normal-ordered Hamiltonian are elements of the Fock matrix, \( F^Q_P \). The anti-symmetrized electron-electron integrals,   $V^{QS}_{PR} = G^{QS}_{PR} -G^{SQ}_{PR}$,   represent the 2e contributions to the normal-ordered Hamiltonian. The summation is formally restricted to pe orbitals but in practice, it is further reduced due to truncations of the occupied and virtual space
commonly employed in correlated calculations. In this form, it is observed that the matrix elements \( F^Q_P \) can be obtained from exact decoupling of the corresponding converged 4c molecular Fock matrix. If the basis chosen for the correlation calculation is taken to be the canonical HF orbitals, the list of nonzero matrix elements reduces to the orbital energies $F^Q_P = \epsilon^+_P \delta_{PQ}$\cite{sikkema2009molecular}. Similarly, we restricted ourselves to the pe 2e interaction terms, $G^{++}_{++}$. These can be obtained for the Coulomb integrals as $G^{++}_{++} = C^{\dagger,+}_{L}C^{\dagger,+}_{L}G^{LL}_{LL}C^{+}_{L}C^{+}_{L} + C^{\dagger,+}_{L}C^{\dagger,+}_{S}G^{LS}_{LS}C^{+}_{L}C^{+}_{S} + C^{\dagger,+}_{S}C^{\dagger,+}_{L}G^{SL}_{SL}C^{+}_{S}C^{+}_{L} + C^{\dagger,+}_{S}C^{\dagger,+}_{S}G^{SS}_{SS}C^{+}_{S}C^{+}_{S}$ and for Gaunt/Breit as $G^{++}_{++} = C^{\dagger,+}_{L}C^{\dagger,+}_{L}G^{SS}_{LL}C^{+}_{S}C^{+}_{S} + C^{\dagger,+}_{L}C^{\dagger,+}_{S}G^{SL}_{LS}C^{+}_{S}C^{+}_{L} + C^{\dagger,+}_{S}C^{\dagger,+}_{L}G^{LS}_{SL}C^{+}_{L}C^{+}_{S} + C^{\dagger,+}_{S}C^{\dagger,+}_{S}G^{LL}_{SS}C^{+}_{L}C^{+}_{L}$\cite{sikkema2009molecular}. All X2C$_{mmf}$ transformed normal-order Hamiltonians used in this study are exact in the 1e Fock term (labeled -1e). X2C$_{mmf}$ transformed normal-order Hamiltonians that are also exact in the 2e interaction terms are labeled -1e2e. and account for all Coulomb integrals ($G^{LL}_{LL}$, $G^{LS}_{LS}$, $G^{SL}_{SL}$, $G^{SS}_{SS}$) and Gaunt/Breit integrals (if used at 4c SCF level, $G^{SS}_{LL}$, $G^{SL}_{LS}$, $G^{LS}_{SL}$, $G^{LL}_{SS}$). Nonexact 2e interaction terms in the X2C$_{mmf}$ transformed normal-order Hamiltonians account only for the transformed Coulomb integrals ($G^{LL}_{LL}$, $G^{LS}_{LS}$, $G^{SL}_{SL}$, $G^{SS}_{SS}$). These variations of X2C$_{mmf}$ transformed normal-order Hamiltonians are defined in Table \ref{tab:x2c}.

\begin{table}[H]
\centering
\begin{tabular}{ccc}
\hline
\textbf{X2C$_{mmf}$ Type} & \textbf{1-electron} & \textbf{2-electron contribution} \\
\hline
DC-1e2e & Coulomb &  Coulomb \\
 DCG-1e & Coulomb-Gaunt&  Coulomb \\
 DCB-1e & Coulomb-Breit & Coulomb\\
 DCG-1e2e & Coulomb-Gaunt & Coulomb-Gaunt\\
 DCB-1e2e & Coulomb-Breit & Coulomb-Breit \\
\hline
\end{tabular}
\caption{X2C$_{mmf}$ transformation used in study as well as their contributions to 1-electron and 2-electron terms.}
\label{tab:x2c}
\end{table}

After X2C$_{mmf}$, Kramers unrestricted CC theory is used to approximate the true wave function by systematically mixing in excited configurations. The wave function in CC theory is defined as $\left| \Psi \right> = e^{\hat{T}} \left| \Phi_{X2C_{mmf},0} \right>$ where $\left| \Phi_{X2C_{mmf},0} \right>$ is the single determinant wave function\cite{shavitt2009many,pototschnig2021implementation}. For this work, the cluster operator $\hat{T}$ was truncated to the SD hole-particle excitations $\hat{T} = \hat{T_1} + \hat{T_2}$ where $\hat{T_1} = \sum_i \sum_a t^i_a \hat{a}^\dagger_a \hat{a}_i$ and $\hat{T_2} = \frac{1}{4} \sum_{ij} \sum_{ab} t^{ij}_{ab} \hat{a}^\dagger_a \hat{a}^\dagger_b \hat{a}_j \hat{a}_i$. Using the similarity transformed Hamiltonian, $\tilde{H} = e^{-\hat{T}}\hat{H}_{X2C_{mmf}}e^{\hat{T}}$, one can solve for the SD excitation amplitudes, $t^i_a$ and $t^{ij}_{ab}$ through various subspace projections listed in Equations \ref{cc1}, \ref{cc2}, and \ref{cc3}\cite{bartlett2012coupled,cizek1969j,purvis1982full,bartlett1989coupled,visscher1995kramers,visscher1996formulation,pototschnig2021implementation},

\begin{equation}
    \left< \Phi_{X2C_{mmf},0} \right| \tilde{H} \left| \Phi_{X2C_{mmf},0} \right> = E,
    \label{cc1}
\end{equation}

\begin{equation}
    \left< \Phi_{X2C_{mmf},0} \right| \hat{a}^\dagger_i \hat{a}_a \tilde{H} \left| \Phi_{X2C_{mmf},0} \right> = 0,
    \label{cc2}
\end{equation}

\begin{equation}
    \left< \Phi_{X2C_{mmf},0} \right| \hat{a}^\dagger_i \hat{a}^\dagger_j \hat{a}_b \hat{a}_a \tilde{H} \left| \Phi_{X2C_{mmf},0} \right> = 0.
    \label{cc3}
\end{equation}

\noindent It is important to note that the asymmetric CCSD equations do not satisfy the variational principle due to the truncation of the cluster operators. This  means that the calculated energy will not necessarily be an upper bound for the exact energy\cite{crawford2007introduction}. The Kramers unrestricted CCSD implementation was benchmarked using PySCF implementation of CCSD. Both the developed CCSD code and PySCF's implementation used the same X2C$_{mmf}$ transformed normal-ordered Hamiltonian as the single determinant reference state. The results of these calculations can be seen in Tables \ref{tab:dhfccsd0}-\ref{tab:dhfccsd3} (convergence tolerance set to 1e$^{-11}$, max DIIS set to 8). From these Tables it is observed that all converged values match to convergence tolerance. Using the CCSD ground-state wave function and noting  that $\tilde{H}$ is not Hermitian, one can define the excited wave function using 
\begin{equation}
\hat{R}_I\left|\Psi \right>= e^{\hat{T}}\left(r_0 + \sum_i \sum_a r^a_i \hat{a}^\dagger_a \hat{a}_i + \frac{1}{4} \sum_{ij} \sum_{ab} r^{ab}_{ij} \hat{a}^\dagger_a \hat{a}^\dagger_b \hat{a}_j \hat{a}_i\right)\left|\Phi_{X2C_{mmf},0} \right>,
\end{equation}

\begin{equation}
\left<\Tilde{\Psi} \right|\hat{L}_I = \left<\Phi_{X2C_{mmf},0} \right|\left(l_0 + \sum_i \sum_a l^i_a \hat{a}^\dagger_a \hat{a}_i + \frac{1}{4} \sum_{ij} \sum_{ab} l^{ij}_{ab} \hat{a}^\dagger_a \hat{a}^\dagger_b \hat{a}_j \hat{a}_i\right)e^{-\hat{T}},
\end{equation}

\noindent where $I$ represents the $I^{th}$ excited state\cite{zhang2024dirac,yuwono2025two,stanton1993equation}. Similar to CCSD, this procedure truncates the operators and assumes that the CCSD ground state captures the dominant dynamical correlations such that the excited states can be represented as excitations on top of the correlated vacuum. These bra and ket states, although not orthonormal among themselves, satisfy biorthogonality  $\left<\Phi_{X2C_{mmf},0} \left|\hat{L}_I e^{-\hat{T}}e^{\hat{T}}\hat{R}_J\right|\Phi_{X2C_{mmf},0} \right> =  C \delta_{IJ}$ and by choosing C to be unity to one can enforce normalization. Defining the normal-ordered similarity transformed Hamiltonian as  $\bar{H} = \tilde{H} - E$ one can solve for the expansion coefficients $r_0$, $r_i^a$, $r_{ij}^{ab}$, $l_0$,$l_a^i$, and $l_{ab}^{ij}$ using Equations \ref{r} and \ref{l},

\begin{equation}
\bar{H}\hat{R}_I\left|\Phi_{X2C_{mmf},0} \right>= \omega_I\hat{R}_I\left|\Phi_{X2C_{mmf},0} \right>,
    \label{r}
\end{equation}

\begin{equation}
\left<\Phi_{X2C_{mmf},0} \right|\hat{L}_I\bar{H}  =\left<\Phi_{X2C_{mmf},0} \right|\hat{L}_I \omega_I,
    \label{l}
\end{equation}

\noindent where $\omega_I$ is the difference in energy between the $I^{th}$ excited state and $E$. In this study, PySCF\cite{sun2018pyscf,sun2020recent} implementation of EOM was used and we refer the reader to their open source code for details of implementation. Using EOM-CCSD-X2C$_{mmf}$-DHF, the ${}^{2}S_{\frac{1}{2}} \rightarrow {}^{2}P_{\frac{1}{2}}$	 and $^{2}S_{\frac{1}{2}} \rightarrow {}^{2}P_{\frac{3}{2}}$ excitations for the alkali group of elements were calculated. Results of these calculations are consistent to experiment  (see in Tables \ref{tab:li_excitation}-\ref{tab:fine_splitting}, convergence tolerance set to 1e$^{-11}$) and a similar study\cite{yuwono2025two} using the Kramers unrestricted EOM-CCSD-X2C$_{mmf}$-DHF, a point-nucleus charge, and the X2C$_{mmf}$ untransformed DC-1e, and DCB-1e Hamiltonian for Na, K, and Rb in the ANO basis sets.

\section{Analysis of Results}
\begin{figure}[H]
    \centering
    \includegraphics[width=.6\textwidth]{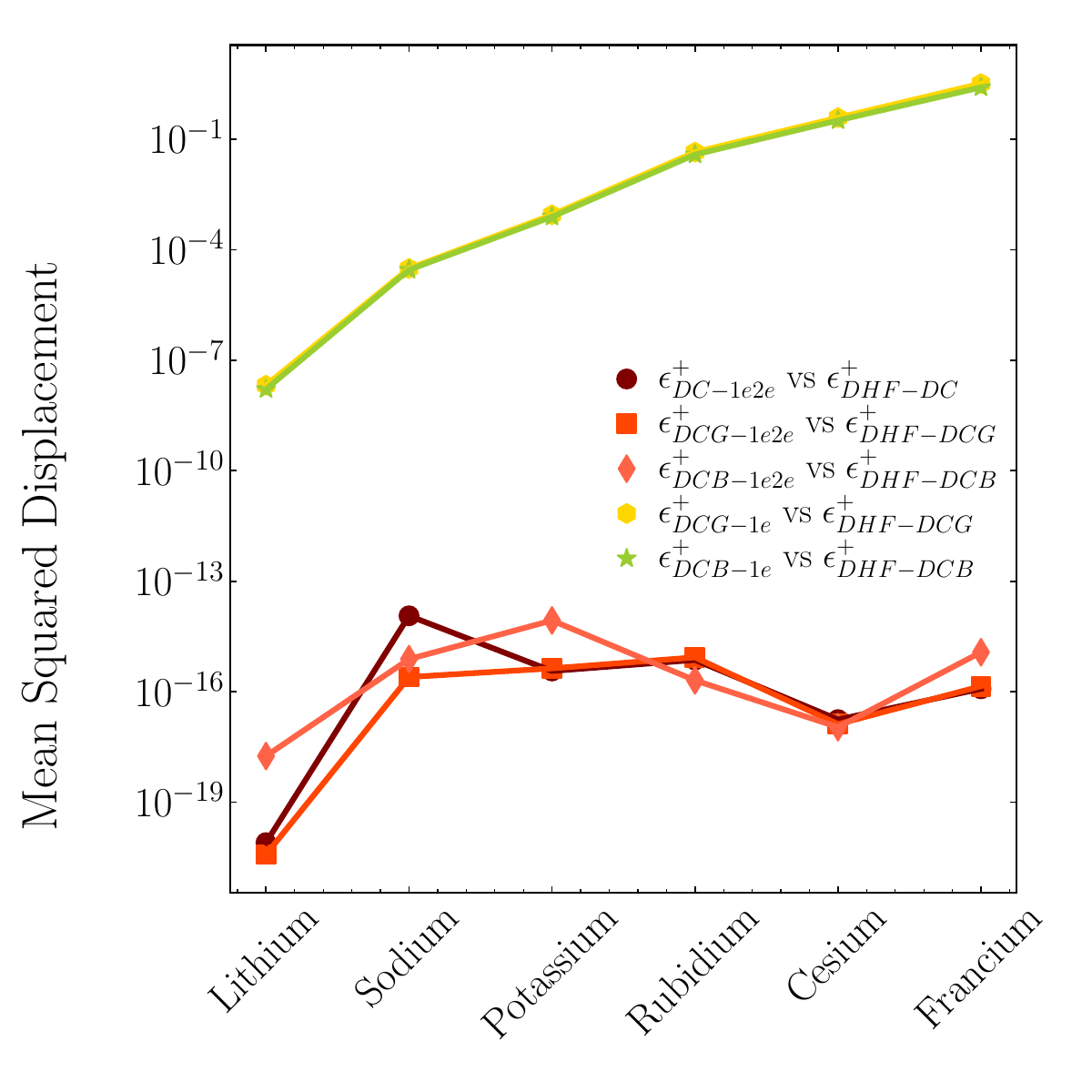}
    \caption{Mean squared displacement of positive energy spectrum after X2C$_{mmf}$ using DC-1e2e, DCG-1e, and DGB-1e, DCG-1e2e, and DCB-1e2e transformation versus the respective positive energy spectrum from 4c DHF using DC, DCG, and DCB Hamiltonian in ANO-RCC-VDZ basis set ($MSD = \frac{1}{n}\sum^{n}_{i=1} |\epsilon^{+}_{i,4c-DHF} - \epsilon^{+}_{i,2c} |^{2}$, units are Hartree).}
    \label{fig:boxplot}
\end{figure}

\begin{figure}[H]
    \centering
    \includegraphics[width=1\textwidth]{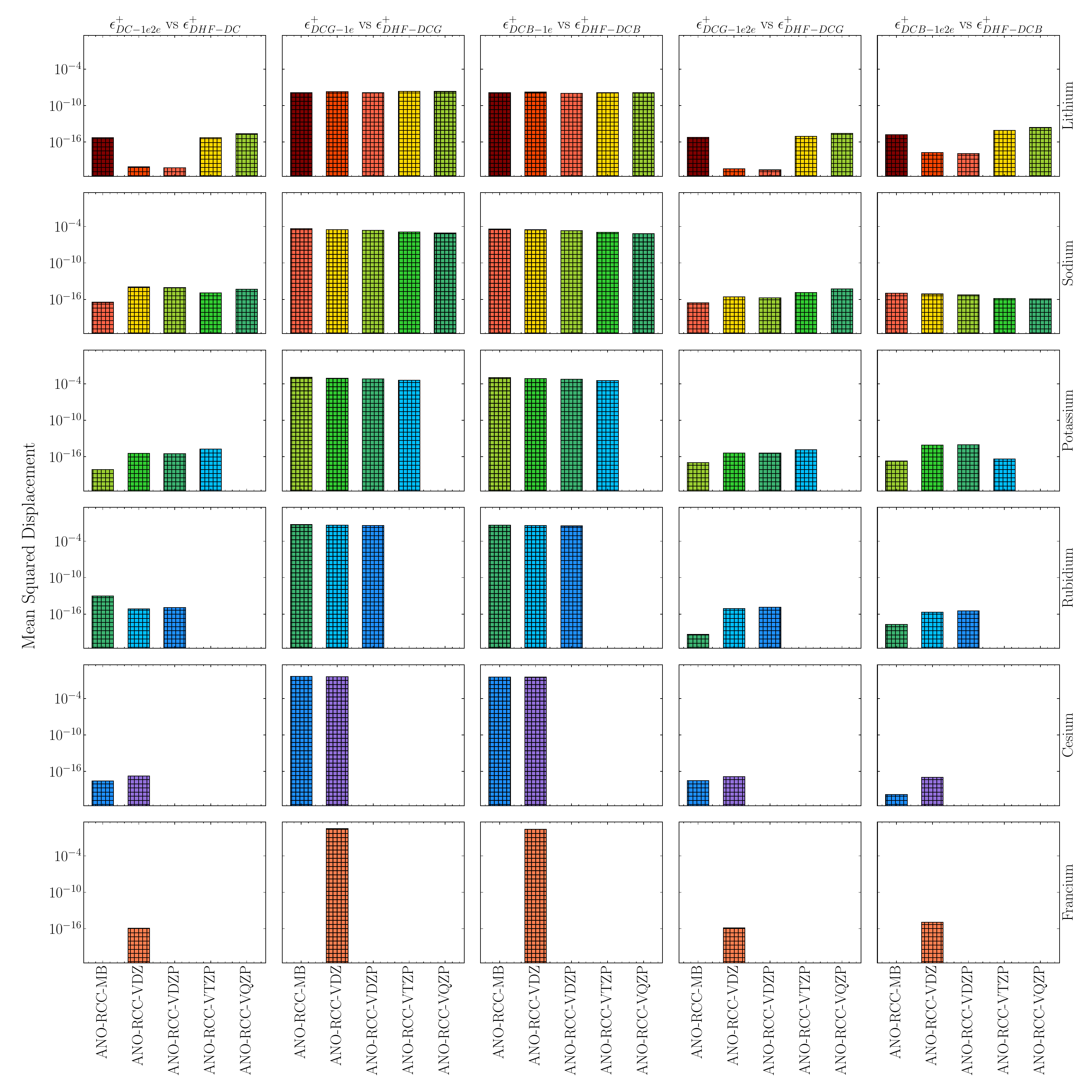}
    \caption{Mean squared displacement of positive energy spectrum after X2C$_{mmf}$ using DC-1e2e, DCG-1e, and DGB-1e, DCG-1e2e, and DCB-1e2e transformation versus the respective positive energy spectrum from 4c DHF using DC, DCG, and DCB Hamiltonian in a variety of basis sets ($MSD = \frac{1}{n}\sum^{n}_{i=1} |\epsilon^{+}_{i,4c-DHF} - \epsilon^{+}_{i,2c} |^{2}$, units are Hartree).}
    \label{fig:boxplot_s}
\end{figure}

The influence of the Gaunt and Breit 2e integrals to their respective DHF-DCG and DHF-DCB mean-fields is quantified using the mean squared displacement of the pe eigenvalue spectrum ($\epsilon^+$) of the 4c Fock (labeled $\epsilon^+_{DHF}$) and decoupled 2c Fock matrix. The $\epsilon^+_{DCG-1e}$, $\epsilon^+_{DCB-1e}$ represent decoupled 2c pe Fock spectrum which are exact in all one-body terms but neglect two-body Gaunt and Breit integrals respectively in decoupling while $\epsilon^+_{DC-1e2e}$, $\epsilon^+_{DCG-1e2e}$, $\epsilon^+_{DCB-1e2e}$ represent the exactly decoupled 2c pe Fock spectrum. Results of these calculations are seen in Figure \ref{fig:boxplot} (Figure \ref{fig:boxplot_s} demonstrates results are independent of basis choice). In this figure, it is observed that the $\epsilon^{+,2c}_{DCG-1e}$ vs $\epsilon^+_{DHF-DCG}$ and $\epsilon^+_{DCB-1e}$ vs $\epsilon^+_{DHF-DCB}$ exhibit a growing discrepancy in the pe eigenvalue spectrum while the exactly decoupled 2c  have consistent negligible error. This confirms that Gaunt and Breit integrals increasingly contribute to their respective 4c DHF mean-field as elements become heavier.

\begin{figure}[H]
    \centering
    \includegraphics[width=.6\textwidth]{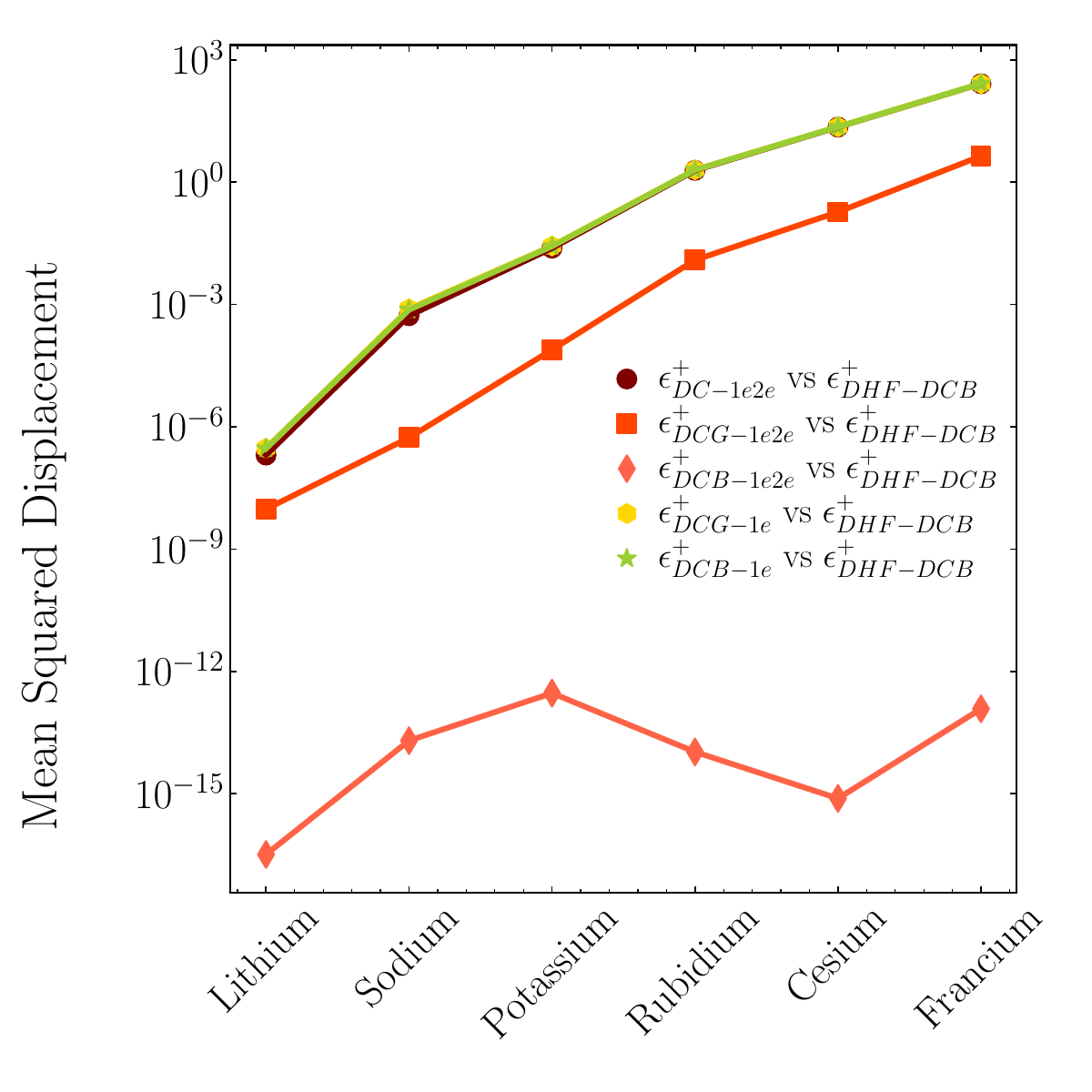}
    \caption{Mean squared displacement of positive energy spectrum after X2C$_{mmf}$ using DC-1e2e, DCG-1e, and DGB-1e, DCG-1e2e, and DCB-1e2e transformation versus the absolute value of positive energy spectrum from 4c DHF using the DCB Hamiltonian in ANO-RCC-VDZ basis set ($MSD = \frac{1}{n}\sum^{n}_{i=1} |\epsilon^{+}_{i,4c-DHF-DCB} - \epsilon^{+}_{i,2c} |^{2}$, units are Hartree).}
    \label{fig:boxplot1}
\end{figure}

\begin{figure}[H]
    \centering
    \includegraphics[width=1\textwidth]{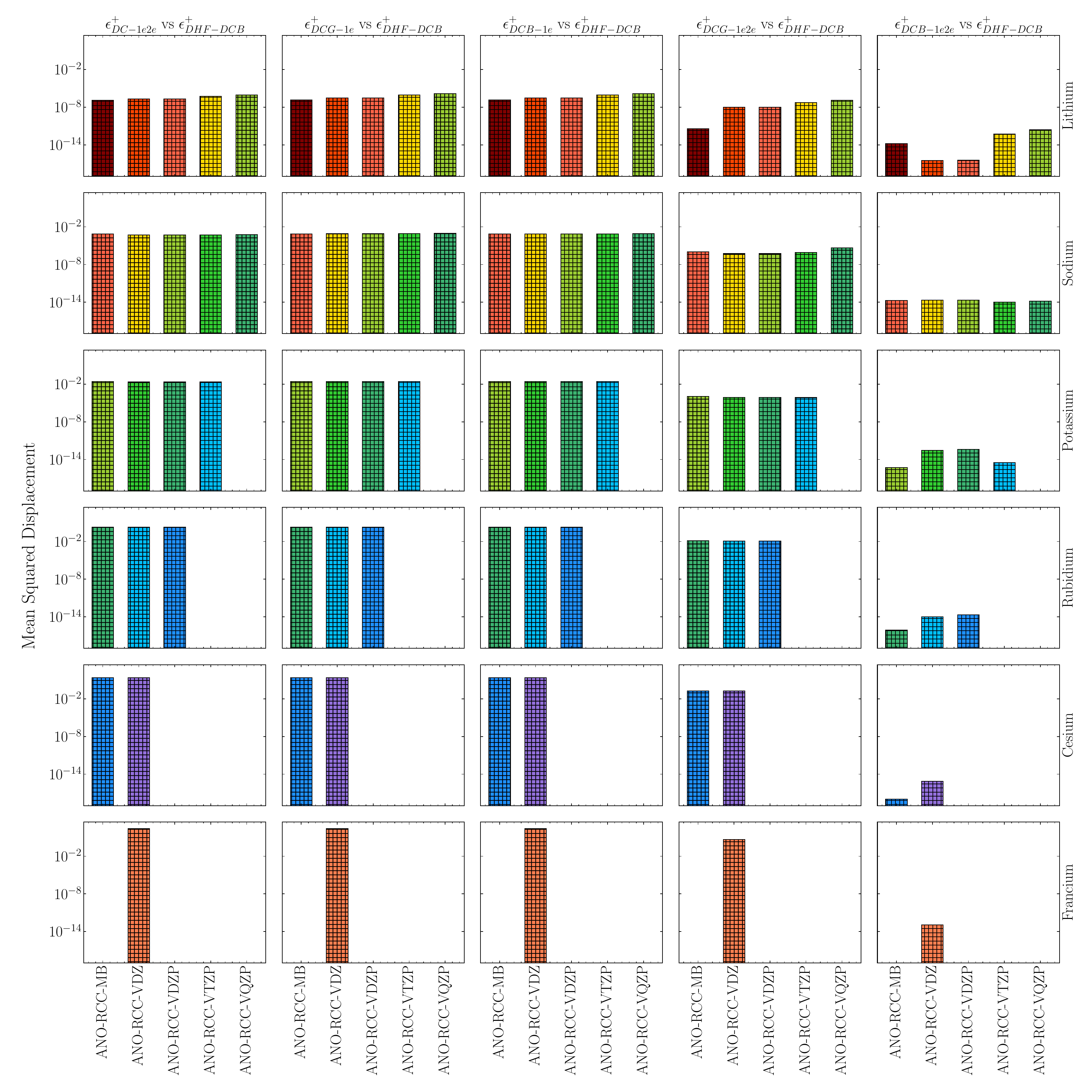}
    \caption{Mean squared displacement of positive energy spectrum after X2C$_{mmf}$ using DC-1e2e, DCG-1e, and DGB-1e, DCG-1e2e, and DCB-1e2e transformation versus the absolute value of positive energy spectrum from 4c DHF using the DCB Hamiltonian in a variety of basis sets ($MSD = \frac{1}{n}\sum^{n}_{i=1} |\epsilon^{+}_{i,4c-DHF-DCB} - \epsilon^{+}_{i,2c} |^{2}$, units are Hartree).}
    \label{fig:boxplot1_s}
\end{figure}

A similar procedure can be performed to quantify any discrepancies in 4c DHF mean fields using higher order relativistic corrections (DC, DCG, and DCB) with increasing atomic number. This is accomplished using the mean squared displacement of the pe eigenvalue spectrum ($\epsilon^+$) of the 4c DHF-DCB Fock (labeled $\epsilon^+_{DHF-DCB}$) and various decoupled 2c Fock introduced above ($\epsilon^+_{DCG-1e}$, $\epsilon^+_{DCB-1e}$, $\epsilon^+_{DC-1e2e}$, $\epsilon^+_{DCG-1e2e}$, $\epsilon^+_{DCB-1e2e}$). Figure \ref{fig:boxplot1} shows the results of mean squared displacement analysis while Figure \ref{fig:boxplot1_s} shows results are independent of basis choice. Observing a growing displacement in pe eigenvalue spectrum neglecting Gaunt or gauge ($\epsilon^+_{DHF-DCB}$ vs $\epsilon^+_{DCG-1e}$, $\epsilon^+_{DCB-1e}$, $\epsilon^+_{DC-1e2e}$, $\epsilon^+_{DCG-1e2e}$), a lower displacement in spectrum exact in Gaunt ($\epsilon^+_{DHF-DCB}$ vs $\epsilon^+_{DCG-1e2e}$), and a negligible displacement in the exactly decoupled spectrum including both Gaunt and gauge ($\epsilon^+_{DHF-DCB}$ vs $\epsilon^+_{DCB-1e2e}$), confirms an increasing discrepancy in pe mean field obtained using 4c DHF-DC, DHF-DCG, or DHF-DCB with increasing atomic number. This result emphasizes the need for increasingly accurate Hamiltonian as atomic number increases.

Moving on to the EOM-CCSD-X2C$_{mmf}$-DHF calculations, it is of interest to understand the impact of the Gaunt and Breit integrals to the 1e and 2e terms of X2C$_{mmf}$ transformed normal-ordered Hamiltonians
used in post 4c SCF correlation-excitations steps. Table \ref{tab:fine_splitting} shows electronic fine structure (EFS, $^{2}P_{\frac{3}{2}} - {}^{2}P_{\frac{1}{2}}$) for varying atom types, basis types, and X2C$_{mmf}$ transformation types, calculated using EOM-CCSD-X2C$_{mmf}$-DHF and compared to experiment. 
Table \ref{tab:fine_splitting} demonstrates that error of the predicted and experimental EFS decreases with increasingly accurate basis sets (ANO-RCC-VDZ $\rightarrow{}$ ANO-RCC-VDZP $\rightarrow{}$ ANO-RCC-VTZP $\rightarrow{}$ ANO-RCC-VQZP). Table \ref{tab:fine_splitting} further demonstrates that as elements increase in atomic number, predictions deviate from experiment. These deviations can likely be attributed the neglect of important higher-order excitations necessary for modeling heavier elements\cite{magoulas2022addressing}. However, for small elements (Na,K), where a CCSD framework is sufficient to provide a reasonable reconstruction of correlation, it is noted that  more accurate X2C$_{mmf}$ transformed normal-ordered Hamiltonians provide a better prediction of the EFS.

\begin{table}[H]
\centering
\footnotesize
\makebox[\textwidth][c]{%
        \begin{tabular}{cccccccc}
\hline
\textbf{Atom} &\textbf{Basis} & \textbf{Experiment}&\textbf{DC-1e2e}& \textbf{DCG-1e} & \textbf{DCB-1e} & \textbf{DCG-1e2e} & \textbf{DCB-1e2e}\\
\hline
Na	&	ANO-RCC-MB	&	0.0021	&	0.0030	&	0.0028	&	0.0028	&	0.0028	&	0.0028	\\
Na	&	ANO-RCC-VDZ	&	0.0021	&	0.0023	&	0.0022	&	0.0022	&	0.0022	&	0.0022	\\
Na	&	ANO-RCC-VDZP	&	0.0021	&	0.0024	&	0.0022	&	0.0022	&	0.0022	&	0.0022	\\
Na	&	ANO-RCC-VTZP	&	0.0021	&	0.0023	&	0.0022	&	0.0022	&	0.0022	&	0.0022	\\
Na	&	ANO-RCC-VQZP	&	0.0021	&	0.0023	&	0.0022	&	0.0022	&	0.0022	&	0.0022	\\
K	&	ANO-RCC-MB	&	0.0072	&	0.0074	&	0.0072	&	0.0072	&	0.0072	&	0.0072	\\
K	&	ANO-RCC-VDZ	&	0.0072	&	0.0066	&	0.0064	&	0.0064	&	0.0064	&	0.0064	\\
K	&	ANO-RCC-VDZP	&	0.0072	&	0.0077	&	0.0075	&	0.0075	&	0.0075	&	0.0075	\\
K	&	ANO-RCC-VTZP	&	0.0072	&	0.0076	&	0.0074	&	0.0074	&	0.0074	&	0.0074	\\
Rb	&	ANO-RCC-MB	&	0.0295	&	0.0220	&	0.0220	&	0.0219	&	0.0220	&	0.0219	\\
Rb	&	ANO-RCC-VDZ	&	0.0295	&	0.0235	&	0.0232	&	0.0232	&	0.0232	&	0.0232	\\
Rb	&	ANO-RCC-VDZP	&	0.0295	&	0.0240	&	0.0237	&	0.0237	&	0.0237	&	0.0237	\\
Cs	&	ANO-RCC-MB	&	0.0687	&	0.0508	&	0.0513	&	0.0512	&	0.0513	&	0.0512	\\
Cs	&	ANO-RCC-VDZ	&	0.0687	&	0.0529	&	0.0529	&	0.0529	&	0.0529	&	0.0529	\\
Fr	&	ANO-RCC-VDZ	&	0.2091	&	0.1543	&	0.1566	&	0.1563	&	0.1566	&	0.1563	\\
\hline
\end{tabular}
}
\caption{Electronic Fine structure splitting ($^{2}P_{\frac{3}{2}} - {}^{2}P_{\frac{1}{2}}$) for varying atom types, basis types, and X2C$_{mmf}$ transformation types, calculated using EOM-CCSD-X2C$_{mmf}$-DHF and compared to experiment (units are eV).}
\label{tab:fine_splitting}
\end{table}

\begin{figure}[H]
    \centering
    \includegraphics[width=.6\textwidth]{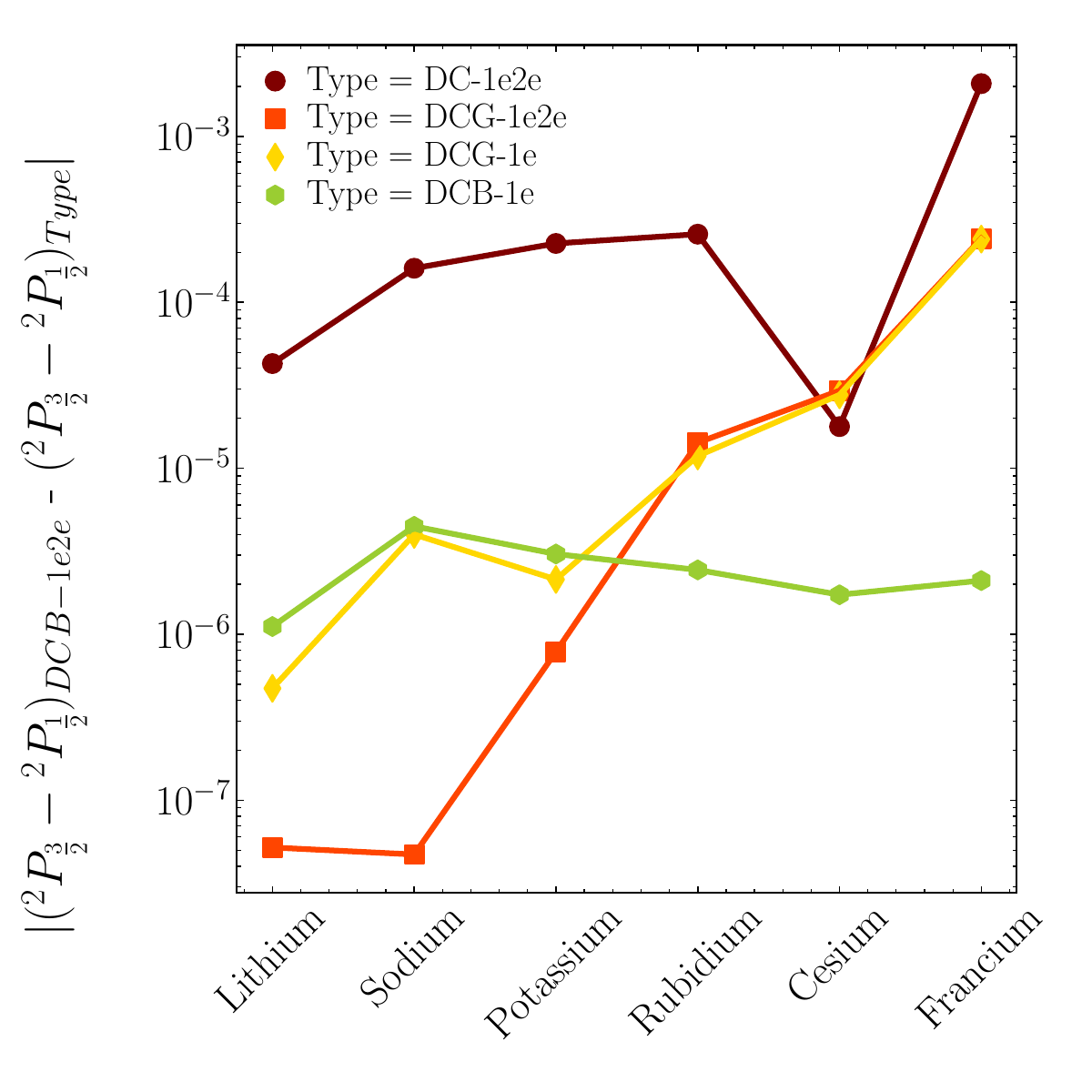}
    \caption{Absolute difference of fine structure splitting predicted using the DC-1e2e, DCG-1e, DCB-1e, and DCG-1e2e transformation versus the DCB-1e2e transformation for various basis sets calculated using EOM-CCSD-X2C$_{mmf}$-DHF in ANO-RCC-VDZ basis set (units are eV).}
    \label{fig:boxplot2}
\end{figure}

\begin{figure}[H]
    \centering
    \includegraphics[width=1\textwidth]{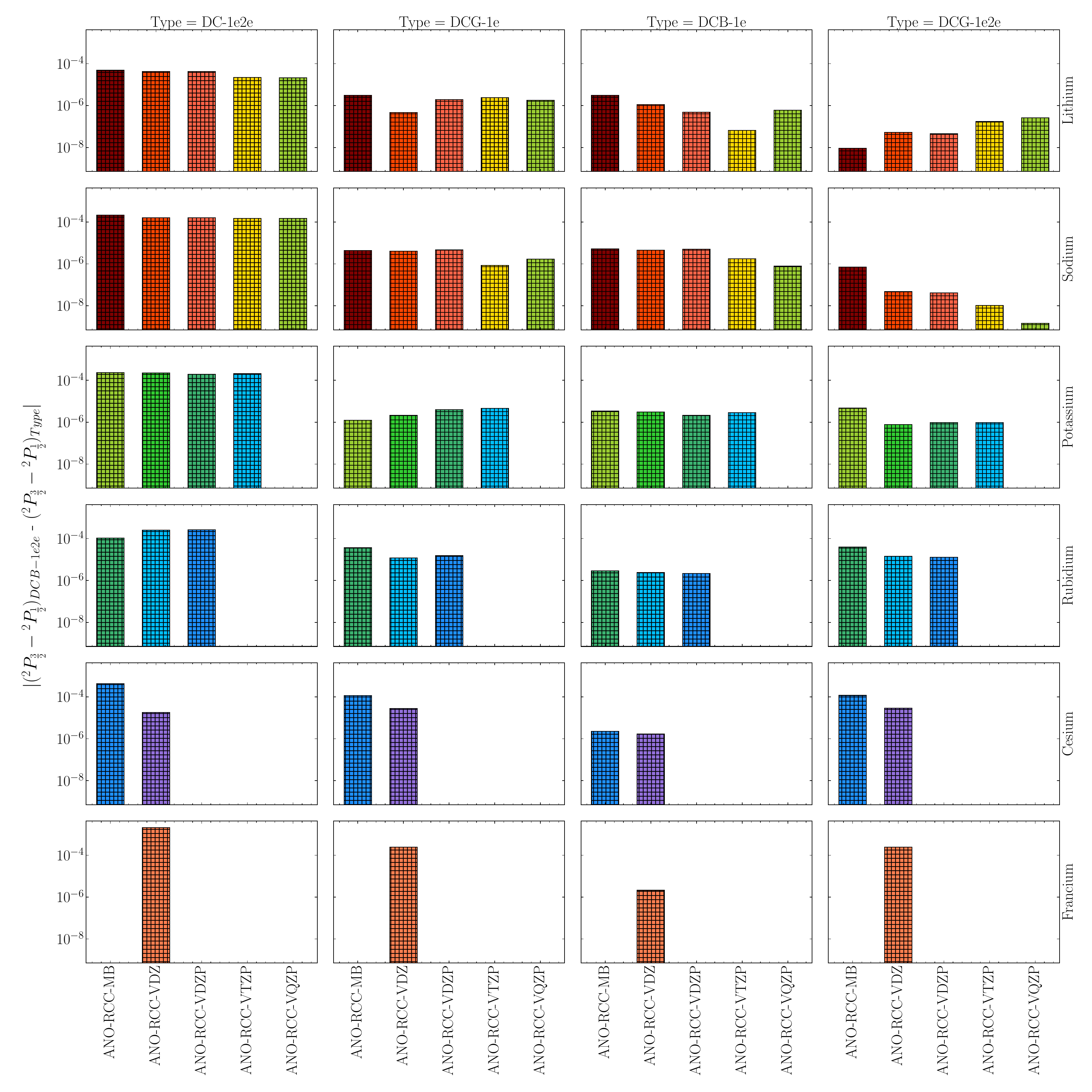}
    \caption{Absolute difference of fine structure splitting predicted using the DC-1e2e, DCG-1e, DCB-1e, and DCG-1e2e transformation versus the DCB-1e2e transformation for various basis sets calculated using EOM-CCSD-X2C$_{mmf}$-DHF in a variety of basis sets (units are eV).}
    \label{fig:boxplot2_s}
\end{figure}

Figure \ref{fig:boxplot2} calculates the absolute difference between the EFS (${}^{2}P_{\frac{3}{2}} - {}^{2}P_{\frac{1}{2}}$ ) predicted by DC-1e2e, DCG-1e, DCB-1e, and DCG-1e2e versus the EFS predicted by DCB-1e2e transformation (Figure \ref{fig:boxplot2_s} demonstrates results are independent of basis choice).  Figure \ref{fig:boxplot2}  and Table \ref{tab:fine_splitting} both indicate that even for elements with small atomic number, including only the Coulomb integrals is insufficient, the Gaunt integrals in both the 1e and 2e term of the X2C$_{mmf}$ transformed normal ordered Hamiltonian contribute in 3$^{rd}$ - 4$^{th}$ decimal of EFS prediction within the EOM-CCSD-X2C$_{mmf}$-DHF framework. Additionally, as elements increase in atomic number, the contributions of the gauge integrals in the 1e and 2e term of the X2C$_{mmf}$ transformed normal ordered Hamiltonian contribute in 4$^{th}$ - 5$^{th}$ decimal of EFS prediction.

\section{Closing Remarks}
As higher fidelity electronic structure is needed, the introduction of relativistic theories becomes increasingly important  to characterize certain systems. In order to address this issue, a code of CCSD was developed using the 4c DHF ground state and  X2C$_{mmf}$ transformed normal-order Hamiltonian incorporating all
1e and 2e contributions from the Coulomb, Gaunt, and Breit operators. Results confirm that Gaunt and Breit integrals increasingly contribute to the DHF solutions causing deviations with increasing atomic number in the mean fields generated with varying orders of relativistic corrections (DC, DCG, and DCB). It is also shown that for elements with small atomic number, the Gaunt integrals in both the 1e and 2e term of the X2C$_{mmf}$ transformed normal ordered Hamiltonian contribute significantly to EFS predictions within the EOM-CCSD-X2C$_{mmf}$-DHF framework and that as elements increase in atomic number, the contributions of the gauge integrals become non-negligible. Overall, this work outlines limitations of various X2C$_{mmf}$ transformations, and lays the ground work for more studies utilizing the DCB Hamiltonian within an exact X2C mean-field approach.

\begin{acknowledgement}
This work was carried out under the auspices of the U.S. Department of Energy (DOE) National Nuclear Security Administration (NNSA) under Contract No. 89233218CNA000001. It was supported by the G.T Seaborg Institute for Transactium Science at Los Alamos National Laboratory (L.M.), the LANL LDRD program (C.L. and R.M.T), and in part by the Center for Integrated Nanotechnologies, a DOE BES user facility, in partnership with the LANL Institutional Computing Program for computational resources.
\end{acknowledgement}

\section{Supplementary Material}
\beginsupplement

\begin{table}[H]
\centering
\tiny
\makebox[\textwidth][c]{%
        \begin{tabular}{cccccccc}
\hline
\textbf{Basis} & \textbf{Excitation} & \textbf{Experiment}&\textbf{DC-1e2e} & \textbf{DCG-1e} & \textbf{DCB-1e} & \textbf{DCG-1e2e} & \textbf{DCB-1e2e}\\
\hline
ANO-RCC-MB	&	$^{2}S_{\frac{1}{2}} \rightarrow {}^{2}P_{\frac{1}{2}}$	&	1.8478	&	1.8467	&	1.8467	&	1.8467	&	1.8467	&	1.8467	\\
	&	$^{2}S_{\frac{1}{2}} \rightarrow {}^{2}P_{\frac{3}{2}}$	&	1.8479	&	1.8468	&	1.8467	&	1.8467	&	1.8467	&	1.8467	\\
ANO-RCC-VDZ	&	$^{2}S_{\frac{1}{2}} \rightarrow {}^{2}P_{\frac{1}{2}}$	&	1.8478	&	1.8437	&	1.8436	&	1.8436	&	1.8436	&	1.8436	\\
	&	$^{2}S_{\frac{1}{2}} \rightarrow {}^{2}P_{\frac{3}{2}}$	&	1.8479	&	1.8438	&	1.8437	&	1.8437	&	1.8437	&	1.8437	\\
ANO-RCC-VDZP	&	$^{2}S_{\frac{1}{2}} \rightarrow {}^{2}P_{\frac{1}{2}}$	&	1.8478	&	1.8369	&	1.8368	&	1.8368	&	1.8368	&	1.8368	\\
	&	$^{2}S_{\frac{1}{2}} \rightarrow {}^{2}P_{\frac{3}{2}}$	&	1.8479	&	1.8370	&	1.8369	&	1.8369	&	1.8369	&	1.8369	\\
ANO-RCC-VTZP	&	$^{2}S_{\frac{1}{2}} \rightarrow {}^{2}P_{\frac{1}{2}}$	&	1.8478	&	1.8434	&	1.8434	&	1.8434	&	1.8434	&	1.8434	\\
	&	$^{2}S_{\frac{1}{2}} \rightarrow {}^{2}P_{\frac{3}{2}}$	&	1.8479	&	1.8436	&	1.8435	&	1.8435	&	1.8435	&	1.8435	\\
ANO-RCC-VQZP	&	$^{2}S_{\frac{1}{2}} \rightarrow {}^{2}P_{\frac{1}{2}}$	&	1.8478	&	1.8455	&	1.8454	&	1.8454	&	1.8454	&	1.8454	\\
	&	$^{2}S_{\frac{1}{2}} \rightarrow {}^{2}P_{\frac{3}{2}}$	&	1.8479	&	1.8456	&	1.8455	&	1.8455	&	1.8456	&	1.8455	\\
\hline
\end{tabular}
}
\caption{Lithium's excitations for varying basis types and X2C$_{mmf}$ trasformation types, calculated using 4c DHF, Kramers unrestricted EOM-CCSD with eV units of energy.}
\label{tab:li_excitation}
\end{table}

\begin{table}[H]
\centering
\tiny
\makebox[\textwidth][c]{%
        \begin{tabular}{cccccccc}
\hline
\textbf{Basis} & \textbf{Excitation} & \textbf{Experiment}&\textbf{DC-1e2e} & \textbf{DCG-1e} & \textbf{DCB-1e} & \textbf{DCG-1e2e} & \textbf{DCB-1e2e}\\
\hline
ANO-RCC-MB	&	$^{2}S_{\frac{1}{2}} \rightarrow {}^{2}P_{\frac{1}{2}}$	&	2.1023	&	2.0198	&	2.0199	&	2.0198	&	2.0199	&	2.0198	\\
	&	$^{2}S_{\frac{1}{2}} \rightarrow {}^{2}P_{\frac{3}{2}}$	&	2.1044	&	2.0228	&	2.0227	&	2.0226	&	2.0227	&	2.0226	\\
ANO-RCC-VDZ	&	$^{2}S_{\frac{1}{2}} \rightarrow {}^{2}P_{\frac{1}{2}}$	&	2.1023	&	1.9891	&	1.9891	&	1.9891	&	1.9891	&	1.9890	\\
	&	$^{2}S_{\frac{1}{2}} \rightarrow {}^{2}P_{\frac{3}{2}}$	&	2.1044	&	1.9914	&	1.9913	&	1.9912	&	1.9913	&	1.9912	\\
ANO-RCC-VDZP	&	$^{2}S_{\frac{1}{2}} \rightarrow {}^{2}P_{\frac{1}{2}}$	&	2.1023	&	1.9870	&	1.9870	&	1.9870	&	1.9870	&	1.9870	\\
	&	$^{2}S_{\frac{1}{2}} \rightarrow {}^{2}P_{\frac{3}{2}}$	&	2.1044	&	1.9894	&	1.9892	&	1.9892	&	1.9892	&	1.9892	\\
ANO-RCC-VTZP	&	$^{2}S_{\frac{1}{2}} \rightarrow {}^{2}P_{\frac{1}{2}}$	&	2.1023	&	2.0460	&	2.0459	&	2.0459	&	2.0459	&	2.0459	\\
	&	$^{2}S_{\frac{1}{2}} \rightarrow {}^{2}P_{\frac{3}{2}}$	&	2.1044	&	2.0483	&	2.0481	&	2.0481	&	2.0481	&	2.0480	\\
ANO-RCC-VQZP	&	$^{2}S_{\frac{1}{2}} \rightarrow {}^{2}P_{\frac{1}{2}}$	&	2.1023	&	2.0755	&	2.0755	&	2.0755	&	2.0755	&	2.0754	\\
	&	$^{2}S_{\frac{1}{2}} \rightarrow {}^{2}P_{\frac{3}{2}}$	&	2.1044	&	2.0778	&	2.0776	&	2.0776	&	2.0777	&	2.0776	\\
\hline
\end{tabular}
}
\caption{Sodium's excitations for varying basis types and X2C$_{mmf}$ trasformation types, calculated using 4c DHF, Kramers unrestricted EOM-CCSD with eV units of energy.}
\label{tab:na_excitation}
\end{table}

\begin{table}[H]
\centering
\tiny
\makebox[\textwidth][c]{%
        \begin{tabular}{cccccccc}
\hline
\textbf{Basis} & \textbf{Excitation} & \textbf{Experiment}&\textbf{DC-1e2e} & \textbf{DCG-1e} & \textbf{DCB-1e} & \textbf{DCG-1e2e} & \textbf{DCB-1e2e}\\
\hline
ANO-RCC-MB	&	$^{2}S_{\frac{1}{2}} \rightarrow {}^{2}P_{\frac{1}{2}}$	&	1.6100	&	1.4465	&	1.4466	&	1.4466	&	1.4466	&	1.4466	\\
	&	$^{2}S_{\frac{1}{2}} \rightarrow {}^{2}P_{\frac{3}{2}}$	&	1.6171	&	1.4539	&	1.4538	&	1.4537	&	1.4538	&	1.4537	\\
ANO-RCC-VDZ	&	$^{2}S_{\frac{1}{2}} \rightarrow {}^{2}P_{\frac{1}{2}}$	&	1.6100	&	1.4467	&	1.4471	&	1.4470	&	1.4471	&	1.4469	\\
	&	$^{2}S_{\frac{1}{2}} \rightarrow {}^{2}P_{\frac{3}{2}}$	&	1.6171	&	1.4533	&	1.4534	&	1.4533	&	1.4534	&	1.4533	\\
ANO-RCC-VDZP	&	$^{2}S_{\frac{1}{2}} \rightarrow {}^{2}P_{\frac{1}{2}}$	&	1.6100	&	1.5678	&	1.5681	&	1.5680	&	1.5681	&	1.5680	\\
	&	$^{2}S_{\frac{1}{2}} \rightarrow {}^{2}P_{\frac{3}{2}}$	&	1.6171	&	1.5755	&	1.5756	&	1.5755	&	1.5756	&	1.5755	\\
ANO-RCC-VTZP	&	$^{2}S_{\frac{1}{2}} \rightarrow {}^{2}P_{\frac{1}{2}}$	&	1.6100	&	1.6275	&	1.6277	&	1.6276	&	1.6277	&	1.6276	\\
	&	$^{2}S_{\frac{1}{2}} \rightarrow {}^{2}P_{\frac{3}{2}}$	&	1.6171	&	1.6352	&	1.6351	&	1.6350	&	1.6351	&	1.6350	\\
\hline
\end{tabular}
}
\caption{Potassium's excitations for varying basis types and X2C$_{mmf}$ trasformation types, calculated using 4c DHF, Kramers unrestricted EOM-CCSD with eV units of energy.}
\label{tab:k_excitation}
\end{table}

\begin{table}[H]
\centering
\tiny
\makebox[\textwidth][c]{%
        \begin{tabular}{cccccccc}
\hline
\textbf{Basis} & \textbf{Excitation} & \textbf{Experiment}&\textbf{DC-1e2e} & \textbf{DCG-1e} & \textbf{DCB-1e} & \textbf{DCG-1e2e} & \textbf{DCB-1e2e}\\
\hline
ANO-RCC-MB	&	$^{2}S_{\frac{1}{2}} \rightarrow {}^{2}P_{\frac{1}{2}}$	&	1.5596	&	1.3455	&	1.3449	&	1.3448	&	1.3448	&	1.3448	\\
	&	$^{2}S_{\frac{1}{2}} \rightarrow {}^{2}P_{\frac{3}{2}}$	&	1.5890	&	1.3675	&	1.3668	&	1.3668	&	1.3668	&	1.3668	\\
ANO-RCC-VDZ	&	$^{2}S_{\frac{1}{2}} \rightarrow {}^{2}P_{\frac{1}{2}}$	&	1.5596	&	1.3665	&	1.3676	&	1.3673	&	1.3676	&	1.3673	\\
	&	$^{2}S_{\frac{1}{2}} \rightarrow {}^{2}P_{\frac{3}{2}}$	&	1.5890	&	1.3900	&	1.3908	&	1.3905	&	1.3908	&	1.3905	\\
ANO-RCC-VDZP	&	$^{2}S_{\frac{1}{2}} \rightarrow {}^{2}P_{\frac{1}{2}}$	&	1.5596	&	1.5174	&	1.5183	&	1.5180	&	1.5183	&	1.5180	\\
	&	$^{2}S_{\frac{1}{2}} \rightarrow {}^{2}P_{\frac{3}{2}}$	&	1.5890	&	1.5414	&	1.5420	&	1.5417	&	1.5420	&	1.5417	\\
\hline
\end{tabular}
}
\caption{Rubidium's excitations for varying basis types and X2C$_{mmf}$ trasformation types, calculated using 4c DHF, Kramers unrestricted EOM-CCSD with eV units of energy.}
\label{tab:rb_excitation}
\end{table}

\begin{table}[H]
\centering
\tiny
\makebox[\textwidth][c]{%
        \begin{tabular}{cccccccc}
\hline
\textbf{Basis} & \textbf{Excitation} & \textbf{Experiment}&\textbf{DC-1e2e} & \textbf{DCG-1e} & \textbf{DCB-1e} & \textbf{DCG-1e2e} & \textbf{DCB-1e2e}\\
\hline
ANO-RCC-MB	&	$^{2}S_{\frac{1}{2}} \rightarrow {}^{2}P_{\frac{1}{2}}$	&	1.3859	&	1.2419	&	1.2420	&	1.2419	&	1.2420	&	1.2419	\\
	&	$^{2}S_{\frac{1}{2}} \rightarrow {}^{2}P_{\frac{3}{2}}$	&	1.4546	&	1.2927	&	1.2934	&	1.2932	&	1.2934	&	1.2932	\\
ANO-RCC-VDZ	&	$^{2}S_{\frac{1}{2}} \rightarrow {}^{2}P_{\frac{1}{2}}$	&	1.3859	&	1.2962	&	1.2988	&	1.2982	&	1.2988	&	1.2982	\\
	&	$^{2}S_{\frac{1}{2}} \rightarrow {}^{2}P_{\frac{3}{2}}$	&	1.4546	&	1.3492	&	1.3517	&	1.3511	&	1.3517	&	1.3511	\\
\hline
\end{tabular}
}
\caption{Cesium's excitations for varying basis types and X2C$_{mmf}$ trasformation types, calculated using 4c DHF, Kramers unrestricted EOM-CCSD with eV units of energy.}
\label{tab:cs_excitation}
\end{table}

\begin{table}[H]
\centering
\tiny
\makebox[\textwidth][c]{%
        \begin{tabular}{cccccccc}
\hline
\textbf{Basis} & \textbf{Excitation} & \textbf{Experiment}&\textbf{DC-1e2e} & \textbf{DCG-1e} & \textbf{DCB-1e} & \textbf{DCG-1e2e} & \textbf{DCB-1e2e}\\
\hline
ANO-RCC-VDZ	&	$^{2}S_{\frac{1}{2}} \rightarrow {}^{2}P_{\frac{1}{2}}$	&	1.5172	&	1.8296	&	1.8407	&	1.8390	&	1.8407	&	1.8390	\\
	&	$^{2}S_{\frac{1}{2}} \rightarrow {}^{2}P_{\frac{3}{2}}$	&	1.7264	&	1.9839	&	1.9973	&	1.9954	&	1.9973	&	1.9954	\\
\hline
\end{tabular}
}
\caption{Francium's excitations for varying basis types and X2C$_{mmf}$ trasformation types, calculated using 4c DHF, Kramers unrestricted EOM-CCSD with eV units of energy.}
\label{tab:fr_excitation}
\end{table}

\begin{table}[H]
\centering
\tiny
\makebox[\textwidth][c]{%
        \begin{tabular}{cccccc}
\hline
\textbf{Atom}& \textbf{X2C$_{mmf}$ Type}& \textbf{Basis} & \textbf{E$^{PySCF}_{DHF}$ [Ha]} & \textbf{$\Delta$ E$_{CCSD}^{This\ Work}$ [Ha]} & \textbf{$\Delta$ E$_{CCSD}^{PySCF}$ [Ha]} \\
\hline
Li	&	DC-1e2e	&	ANO-RCC-MB	&	-7.43337869	&	-4.58212741E-05	&	-4.58212741E-05	\\
Li	&	DC-1e2e	&	ANO-RCC-VDZ	&	-7.43349063	&	-1.61383218E-02	&	-1.61383218E-02	\\
Li	&	DC-1e2e	&	ANO-RCC-VDZP	&	-7.43349063	&	-1.61401491E-02	&	-1.61401491E-02	\\
Li	&	DC-1e2e	&	ANO-RCC-VTZP	&	-7.43350415	&	-3.65209750E-02	&	-3.65209750E-02	\\
Li	&	DC-1e2e	&	ANO-RCC-VQZP	&	-7.43351115	&	-3.90337186E-02	&	-3.90337186E-02	\\
Li	&	DCG-1e	&	ANO-RCC-MB	&	-7.43337869	&	-4.58212741E-05	&	-4.58212741E-05	\\
Li	&	DCG-1e	&	ANO-RCC-VDZ	&	-7.43349063	&	-1.61383218E-02	&	-1.61383218E-02	\\
Li	&	DCG-1e	&	ANO-RCC-VDZP	&	-7.43349063	&	-1.61401491E-02	&	-1.61401491E-02	\\
Li	&	DCG-1e	&	ANO-RCC-VTZP	&	-7.43350415	&	-3.65209750E-02	&	-3.65209750E-02	\\
Li	&	DCG-1e	&	ANO-RCC-VQZP	&	-7.43351115	&	-3.90337186E-02	&	-3.90337186E-02	\\
Li	&	DCB-1e	&	ANO-RCC-MB	&	-7.43337869	&	-4.58212741E-05	&	-4.58212741E-05	\\
Li	&	DCB-1e	&	ANO-RCC-VDZ	&	-7.43349063	&	-1.61383218E-02	&	-1.61383218E-02	\\
Li	&	DCB-1e	&	ANO-RCC-VDZP	&	-7.43349063	&	-1.61401491E-02	&	-1.61401491E-02	\\
Li	&	DCB-1e	&	ANO-RCC-VTZP	&	-7.43350415	&	-3.65209750E-02	&	-3.65209750E-02	\\
Li	&	DCB-1e	&	ANO-RCC-VQZP	&	-7.43351115	&	-3.90337186E-02	&	-3.90337186E-02	\\
Li	&	DCG-1e2e	&	ANO-RCC-MB	&	-7.43337869	&	-4.58212741E-05	&	-4.58212741E-05	\\
Li	&	DCG-1e2e	&	ANO-RCC-VDZ	&	-7.43349063	&	-1.61383218E-02	&	-1.61383218E-02	\\
Li	&	DCG-1e2e	&	ANO-RCC-VDZP	&	-7.43349063	&	-1.61401491E-02	&	-1.61401491E-02	\\
Li	&	DCG-1e2e	&	ANO-RCC-VTZP	&	-7.43350415	&	-3.65209750E-02	&	-3.65209750E-02	\\
Li	&	DCB-1e2e	&	ANO-RCC-MB	&	-7.43337869	&	-4.58212741E-05	&	-4.58212741E-05	\\
Li	&	DCB-1e2e	&	ANO-RCC-VDZ	&	-7.43349063	&	-1.61383218E-02	&	-1.61383218E-02	\\
Li	&	DCB-1e2e	&	ANO-RCC-VDZP	&	-7.43349063	&	-1.61401491E-02	&	-1.61401491E-02	\\
Li	&	DCB-1e2e	&	ANO-RCC-VTZP	&	-7.43350415	&	-3.65209750E-02	&	-3.65209750E-02	\\
Li	&	DCB-1e2e	&	ANO-RCC-VQZP	&	-7.43351115	&	-3.90337186E-02	&	-3.90337186E-02	\\
\hline
\end{tabular}
}
\caption{Tabulated atom type, X2C$_{mmf}$ transformation type, basis, PySCF converged 4c DHF energy, converged unrestricted Kramers CCSD energy correction, and converged PySCF CCSD energy correction.}
\label{tab:dhfccsd0}
\end{table}

\begin{table}[H]
\centering
\tiny
\makebox[\textwidth][c]{%
        \begin{tabular}{cccccc}
\hline
\textbf{Atom}& \textbf{X2C$_{mmf}$ Type}& \textbf{Basis} & \textbf{E$^{PySCF}_{DHF}$ [Ha]} & \textbf{$\Delta$ E$_{CCSD}^{This\ Work}$ [Ha]} & \textbf{$\Delta$ E$_{CCSD}^{PySCF}$ [Ha]} \\
\hline
Na	&	DC-1e2e	&	ANO-RCC-MB	&	-162.05715650	&	-2.20078705E-04	&	-2.20078709E-04	\\
Na	&	DC-1e2e	&	ANO-RCC-VDZ	&	-162.05720863	&	-1.17564643E-01	&	-1.17564643E-01	\\
Na	&	DC-1e2e	&	ANO-RCC-VDZP	&	-162.05720863	&	-1.18638412E-01	&	-1.18638412E-01	\\
Na	&	DC-1e2e	&	ANO-RCC-VTZP	&	-162.05723287	&	-2.03866922E-01	&	-2.03866922E-01	\\
Na	&	DC-1e2e	&	ANO-RCC-VQZP	&	-162.05726902	&	-2.40431049E-01	&	-2.40431049E-01	\\
Na	&	DCG-1e	&	ANO-RCC-MB	&	-162.05715650	&	-2.20078705E-04	&	-2.20078709E-04	\\
Na	&	DCG-1e	&	ANO-RCC-VDZ	&	-162.05720863	&	-1.17564643E-01	&	-1.17564643E-01	\\
Na	&	DCG-1e	&	ANO-RCC-VDZP	&	-162.05720863	&	-1.18638412E-01	&	-1.18638412E-01	\\
Na	&	DCG-1e	&	ANO-RCC-VTZP	&	-162.05723287	&	-2.03866922E-01	&	-2.03866922E-01	\\
Na	&	DCG-1e	&	ANO-RCC-VQZP	&	-162.05726902	&	-2.40431049E-01	&	-2.40431049E-01	\\
Na	&	DCB-1e	&	ANO-RCC-MB	&	-162.05715650	&	-2.20078705E-04	&	-2.20078709E-04	\\
Na	&	DCB-1e	&	ANO-RCC-VDZ	&	-162.05720863	&	-1.17564643E-01	&	-1.17564643E-01	\\
Na	&	DCB-1e	&	ANO-RCC-VDZP	&	-162.05720863	&	-1.18638412E-01	&	-1.18638412E-01	\\
Na	&	DCB-1e	&	ANO-RCC-VTZP	&	-162.05723287	&	-2.03866922E-01	&	-2.03866922E-01	\\
Na	&	DCB-1e	&	ANO-RCC-VQZP	&	-162.05726902	&	-2.40431049E-01	&	-2.40431049E-01	\\
Na	&	DCG-1e2e	&	ANO-RCC-MB	&	-162.05715650	&	-2.20078705E-04	&	-2.20078709E-04	\\
Na	&	DCG-1e2e	&	ANO-RCC-VDZ	&	-162.05720863	&	-1.17564643E-01	&	-1.17564643E-01	\\
Na	&	DCG-1e2e	&	ANO-RCC-VDZP	&	-162.05720863	&	-1.18638412E-01	&	-1.18638412E-01	\\
Na	&	DCG-1e2e	&	ANO-RCC-VTZP	&	-162.05723287	&	-2.03866922E-01	&	-2.03866922E-01	\\
Na	&	DCG-1e2e	&	ANO-RCC-VQZP	&	-162.05726902	&	-2.40431049E-01	&	-2.40431049E-01	\\
Na	&	DCB-1e2e	&	ANO-RCC-MB	&	-162.05715650	&	-2.20078705E-04	&	-2.20078709E-04	\\
Na	&	DCB-1e2e	&	ANO-RCC-VDZ	&	-162.05720863	&	-1.17564643E-01	&	-1.17564643E-01	\\
Na	&	DCB-1e2e	&	ANO-RCC-VDZP	&	-162.05720863	&	-1.18638412E-01	&	-1.18638412E-01	\\
Na	&	DCB-1e2e	&	ANO-RCC-VTZP	&	-162.05723287	&	-2.03866922E-01	&	-2.03866922E-01	\\
Na	&	DCB-1e2e	&	ANO-RCC-VQZP	&	-162.05726902	&	-2.40431049E-01	&	-2.40431049E-01	\\
\hline
\end{tabular}
}
\caption{Tabulated atom type, X2C$_{mmf}$ transformation type, basis, PySCF converged 4c DHF energy, converged unrestricted Kramers CCSD energy correction, and converged PySCF CCSD energy correction.}
\label{tab:dhfccsd1}
\end{table}

\begin{table}[H]
\centering
\tiny
\makebox[\textwidth][c]{%
        \begin{tabular}{cccccc}
\hline
\textbf{Atom}& \textbf{X2C$_{mmf}$ Type}& \textbf{Basis} & \textbf{E$^{PySCF}_{DHF}$ [Ha]} & \textbf{$\Delta$ E$_{CCSD}^{This\ Work}$ [Ha]} & \textbf{$\Delta$ E$_{CCSD}^{PySCF}$ [Ha]} \\
\hline
K	&	DC-1e2e	&	ANO-RCC-MB	&	-601.33547064	&	-2.81712233E-04	&	-2.81712226E-04	\\
K	&	DC-1e2e	&	ANO-RCC-VDZ	&	-601.33574000	&	-4.19427003E-02	&	-4.19427003E-02	\\
K	&	DC-1e2e	&	ANO-RCC-VDZP	&	-601.33574000	&	-1.88640717E-01	&	-1.88640717E-01	\\
K	&	DC-1e2e	&	ANO-RCC-VTZP	&	-601.33586043	&	-2.60677803E-01	&	-2.60677804E-01	\\
K	&	DCG-1e	&	ANO-RCC-MB	&	-601.33547064	&	-2.81712233E-04	&	-2.81712226E-04	\\
K	&	DCG-1e	&	ANO-RCC-VDZ	&	-601.33574000	&	-4.19427003E-02	&	-4.19427003E-02	\\
K	&	DCG-1e	&	ANO-RCC-VDZP	&	-601.33574000	&	-1.88640717E-01	&	-1.88640717E-01	\\
K	&	DCG-1e	&	ANO-RCC-VTZP	&	-601.33586043	&	-2.60677803E-01	&	-2.60677804E-01	\\
K	&	DCB-1e	&	ANO-RCC-MB	&	-601.33547064	&	-2.81712233E-04	&	-2.81712226E-04	\\
K	&	DCB-1e	&	ANO-RCC-VDZ	&	-601.33574000	&	-4.19427003E-02	&	-4.19427003E-02	\\
K	&	DCB-1e	&	ANO-RCC-VDZP	&	-601.33574000	&	-1.88640717E-01	&	-1.88640717E-01	\\
K	&	DCB-1e	&	ANO-RCC-VTZP	&	-601.33586043	&	-2.60677803E-01	&	-2.60677804E-01	\\
K	&	DCG-1e2e	&	ANO-RCC-MB	&	-601.33547064	&	-2.81712233E-04	&	-2.81712226E-04	\\
K	&	DCG-1e2e	&	ANO-RCC-VDZ	&	-601.33574000	&	-4.19427003E-02	&	-4.19427003E-02	\\
K	&	DCG-1e2e	&	ANO-RCC-VDZP	&	-601.33574000	&	-1.88640717E-01	&	-1.88640717E-01	\\
K	&	DCG-1e2e	&	ANO-RCC-VTZP	&	-601.33586043	&	-2.60677803E-01	&	-2.60677804E-01	\\
K	&	DCB-1e2e	&	ANO-RCC-MB	&	-601.33547064	&	-2.81712233E-04	&	-2.81712226E-04	\\
K	&	DCB-1e2e	&	ANO-RCC-VDZ	&	-601.33574000	&	-4.19427003E-02	&	-4.19427003E-02	\\
K	&	DCB-1e2e	&	ANO-RCC-VDZP	&	-601.33574000	&	-1.88640717E-01	&	-1.88640717E-01	\\
K	&	DCB-1e2e	&	ANO-RCC-VTZP	&	-601.33586043	&	-2.60677803E-01	&	-2.60677804E-01	\\

\hline
\end{tabular}
}
\caption{Tabulated atom type, X2C$_{mmf}$ transformation type, basis, PySCF converged 4c DHF energy, converged unrestricted Kramers CCSD energy correction, and converged PySCF CCSD energy correction.}
\label{tab:dhfccsd2}
\end{table}

\begin{table}[H]
\centering
\tiny
\makebox[\textwidth][c]{%
        \begin{tabular}{cccccc}
\hline
\textbf{Atom}& \textbf{X2C$_{mmf}$ Type}& \textbf{Basis} & \textbf{E$^{PySCF}_{DHF}$ [Ha]} & \textbf{$\Delta$ E$_{CCSD}^{This\ Work}$ [Ha]} & \textbf{$\Delta$ E$_{CCSD}^{PySCF}$ [Ha]} \\
\hline
Rb	&	DC-1e2e	&	ANO-RCC-MB	&	-2989.57808707	&	-2.54348217E-04	&	-2.54348209E-04	\\
Rb	&	DC-1e2e	&	ANO-RCC-VDZ	&	-2989.58340243	&	-2.93237276E-02	&	-2.93237276E-02	\\
Rb	&	DC-1e2e	&	ANO-RCC-VDZP	&	-2989.58353577	&	-1.65029862E-01	&	-1.65029862E-01	\\
Rb	&	DCG-1e	&	ANO-RCC-MB	&	-2989.57808707	&	-2.54348217E-04	&	-2.54348209E-04	\\
Rb	&	DCG-1e	&	ANO-RCC-VDZ	&	-2989.58340243	&	-2.93237276E-02	&	-2.93237276E-02	\\
Rb	&	DCG-1e	&	ANO-RCC-VDZP	&	-2989.58353577	&	-1.65029862E-01	&	-1.65029862E-01	\\
Rb	&	DCB-1e	&	ANO-RCC-MB	&	-2989.57808707	&	-2.54348217E-04	&	-2.54348209E-04	\\
Rb	&	DCB-1e	&	ANO-RCC-VDZ	&	-2989.58340243	&	-2.93237276E-02	&	-2.93237276E-02	\\
Rb	&	DCB-1e	&	ANO-RCC-VDZP	&	-2989.58353577	&	-1.65029862E-01	&	-1.65029862E-01	\\
Rb	&	DCG-1e2e	&	ANO-RCC-MB	&	-2989.57808707	&	-2.54348217E-04	&	-2.54348209E-04	\\
Rb	&	DCG-1e2e	&	ANO-RCC-VDZ	&	-2989.58340243	&	-2.93237276E-02	&	-2.93237276E-02	\\
Rb	&	DCG-1e2e	&	ANO-RCC-VDZP	&	-2989.58353577	&	-1.65029862E-01	&	-1.65029862E-01	\\
Rb	&	DCB-1e2e	&	ANO-RCC-MB	&	-2989.57808707	&	-2.54348217E-04	&	-2.54348209E-04	\\
Rb	&	DCB-1e2e	&	ANO-RCC-VDZ	&	-2989.58340243	&	-2.93237276E-02	&	-2.93237276E-02	\\
Rb	&	DCB-1e2e	&	ANO-RCC-VDZP	&	-2989.58353577	&	-1.65029862E-01	&	-1.65029862E-01	\\
Cs	&	DC-1e2e	&	ANO-RCC-MB	&	-8051.50915614	&	-2.97025801E-04	&	-2.97025799E-04	\\
Cs	&	DC-1e2e	&	ANO-RCC-VDZ	&	-8051.57094011	&	-2.52678284E-02	&	-2.52678285E-02	\\
Cs	&	DCG-1e	&	ANO-RCC-MB	&	-8051.50915614	&	-2.97025801E-04	&	-2.97025799E-04	\\
Cs	&	DCG-1e	&	ANO-RCC-VDZ	&	-8051.57094011	&	-2.52678284E-02	&	-2.52678285E-02	\\
Cs	&	DCB-1e	&	ANO-RCC-MB	&	-8051.50915614	&	-2.97025801E-04	&	-2.97025799E-04	\\
Cs	&	DCB-1e	&	ANO-RCC-VDZ	&	-8051.57094011	&	-2.52678284E-02	&	-2.52678285E-02	\\
Cs	&	DCG-1e2e	&	ANO-RCC-MB	&	-8051.50915614	&	-2.97025801E-04	&	-2.97025799E-04	\\
Cs	&	DCG-1e2e	&	ANO-RCC-VDZ	&	-8051.57094011	&	-2.52678284E-02	&	-2.52678285E-02	\\
Cs	&	DCB-1e2e	&	ANO-RCC-MB	&	-8051.50915614	&	-2.97025801E-04	&	-2.97025799E-04	\\
Cs	&	DCB-1e2e	&	ANO-RCC-VDZ	&	-8051.57094011	&	-2.52678284E-02	&	-2.52678285E-02	\\
Fr	&	DC-1e2e	&	ANO-RCC-VDZ	&	-28955.29686485	&	-2.81453993E-02	&	-	\\
Fr	&	DCG-1e	&	ANO-RCC-VDZ	&	-28955.29686485	&	-2.81453993E-02	&	-	\\
Fr	&	DCB-1e	&	ANO-RCC-VDZ	&	-28955.29686485	&	-2.81453993E-02	&	-	\\
Fr	&	DCG-1e2e	&	ANO-RCC-VDZ	&	-28955.29686485	&	-2.81453993E-02	&	-	\\
Fr	&	DCB-1e2e	&	ANO-RCC-VDZ	&	-28955.29686485	&	-2.81453993E-02	&	-	\\

\hline
\end{tabular}
}
\caption{Tabulated atom type, X2C$_{mmf}$ transformation type, basis, PySCF converged 4c DHF energy, converged unrestricted Kramers CCSD energy correction, and converged PySCF CCSD energy correction.}
\label{tab:dhfccsd3}
\end{table}

\begin{table}[H]
\centering
\tiny
\makebox[\textwidth][c]{%
        \begin{tabular}{cccc}
\hline
\textbf{Atom} & \textbf{Basis} & \textbf{X2C$_{mmf}$ Type} & \textbf{Transformation Time [sec]}\\
\hline
Na & ANO-RCC-VDZ & DC-1e2e & 1.2531 \\
Na & ANO-RCC-VDZ & DCG-1e & 1.2207 \\
Na & ANO-RCC-VDZ & DCB-1e & 1.2106 \\
Na & ANO-RCC-VDZ & DCG-1e2e & 1.9817 \\
Na & ANO-RCC-VDZ & DCB-1e2e & 10.2324 \\
Na & ANO-RCC-VDZP & DC-1e2e & 2.6609 \\
Na & ANO-RCC-VDZP & DCG-1e & 2.76 \\
Na & ANO-RCC-VDZP & DCB-1e & 2.8776 \\
Na & ANO-RCC-VDZP & DCG-1e2e & 4.4292 \\
Na & ANO-RCC-VDZP & DCB-1e2e & 23.662 \\
Na & ANO-RCC-VTZP & DC-1e2e & 11.5411  \\
Na & ANO-RCC-VTZP & DCG-1e & 11.555  \\
Na & ANO-RCC-VTZP & DCB-1e & 11.5175  \\
Na & ANO-RCC-VTZP & DCG-1e2e & 20.0802  \\
Na & ANO-RCC-VTZP & DCB-1e2e & 64.6176  \\
Na & ANO-RCC-VQZP & DC-1e2e & 55.9794   \\
Na & ANO-RCC-VQZP & DCG-1e & 56.7785   \\
Na & ANO-RCC-VQZP & DCB-1e & 56.455    \\
Na & ANO-RCC-VQZP & DCG-1e2e & 113.3471    \\
Na & ANO-RCC-VQZP & DCB-1e2e & 180.4449    \\
\hline

\end{tabular}
}
\caption{Time required to perform a variety of X2C$_{mmf}$ transformation  for a variety of basis for a Sodium atom.}
\label{tab:x2c_time}
\end{table}


\bibliography{achemso-demo}

\end{document}